\documentclass{WileyMSP-template}

\usepackage{siunitx} 
\usepackage{caption}
\usepackage{subcaption}
\usepackage{floatrow}
\usepackage[utf8]{inputenc} 
\usepackage{array}
\usepackage{makecell}
\usepackage[table]{xcolor}
\usepackage{hyperref}

\newfloatcommand{capbtabbox}{table}[][\FBwidth]

\begin{document}

\pagestyle{fancy}
\rhead{}

\title{Towards Neural-Network-based optical temperature sensing of Semiconductor Membrane External Cavity Laser}

\maketitle


\author{Jakob Mannstadt}
\author{Arash Rahimi-Iman*}


\dedication{}

\begin{affiliations}
Jakob Mannstadt\\
I. Physikalisches Institut and Center for Materials Research, 
Justus-Liebig-Universität Gießen, 35392 Gießen, Germany  \\

Arash Rahimi-Iman\\
I. Physikalisches Institut and Center for Materials Research, 
Justus-Liebig-Universität Gießen, 35392 Gießen, Germany  \\
*E-mail: arash.rahimi-iman@physik.jlug.de

\end{affiliations}



\medskip

\begin{abstract}
A machine-learning non-contact method to determine the temperature of a laser gain medium via its laser emission with a trained few-layer neural net model is presented. The training of the feed-forward Neural Network (NN) enables the prediction of the device's properties solely from spectral data, here recorded by visible-/nearinfrared-light compact micro-spectrometers for both a diode pump laser and optically-pumped gain membrane of a semiconductor disk laser. Fiber spectrometers are used for the acquisition of large quantities of labelled intensity data, which can afterwards be used for the prediction process. Such pretrained deep NNs enable a fast, reliable and easy way to infer the temperature of a laser system such as our Membrane External Cavity Laser, at a later monitoring stage without the need of additional optical diagnostics or read-out temperature sensors. With the miniature mobile spectrometer and the remote detection ability, the temperature inference capability can be adapted for various laser diodes using transfer learning methods with pretrained models. Here, mean-square-error values for the temperature inference corresponding to sub-percent accuracy of our sensor scheme are reached, while computational cost can be saved by reducing the network depth at the here displayed cost of accuracy, as appropriate for different application scenarios.
\end{abstract}


\section{Introduction}
The development and the design of various optical sensors has facilitated the monitoring of chemical compositions \cite{wolfbeis2006fiber}, the human physiology \cite{vavrinski:2022}, temperature in technical installations \cite{Ashry:22} and detection of anomalies e.g. caused by fire \cite{khan:2022}, to name but a few. 

\smallskip

Optical temperature sensors often rely on the reflection, propagation or interference of light that
directly shines onto the object of which the temperature shall be determined. 
A number of different techniques are known in this domain, such as those using distributed Bragg gratings, Fabry-Perot microcavities or Raman and Rayleigh scattering \cite{lee:2007,rai:2007}.
Fiber Bragg Gratings are based on the wavelength shift of the peak reflectivity due to temperature-dependent refractive index changes. Consequently, the change of the wavelength can be employed as a temperature measure.
In Fabry Perot Interferometer the change of the relative distance of the mirrors due to the thermal expansion is employed. Thus, the peak transmission wavelength indicates the temperature in this type of sensor.
Sensing methods that rely on direct backscattering of light from the target for instance are based on Raman scattering, where the ratio of anti-Stokes and Stokes scattered light powers is temperature dependent \cite{rai:2007}.

\smallskip

In contrast, regarding light emitter devices, one can directly access thermal shifts for sensing properties. The quantum defect and other LASER device or LED inefficiencies will heat up the optical emitter system, which can cause a red shift on the emission wavelength, e.g. of the laser (gain spectrum). Such red shift can be employed to deduce the temperature of the emitter system. In semiconductor lasers, it arises mainly due to shift in the gain spectrum of the quantum well (QW) induced by a change in the band gap of the gain material. In addition, a rise in temperature thermally expands the typically employed microcavity of the laser system, generating another red shift of the emission wavelength. However, the change due to the gain spectrum of the QW dominates over the thermal expansion of the cavity \cite{heinen_temperature_gain, bedford}. The resulting overall output modification can be a suitable indicator of the device temperature.

\smallskip

To analyse the optical response, i.e. the light signal alterations according to the sensing principle, complex or expensive tools are typically employed, such as to probe wavelength shifts, pulse properties, signal levels, and so forth. Thus, they require some form of specialized detector or other additional devices to fulfil one or more tasks.

\smallskip

In contrast to aforementioned methods, techniques can be used which deduce temperature changes in an active medium by modifications of its irradiation properties \cite{patent_china, patent_eu}. Here, instead of having a mechanical or electronic sensor head in touch with the emitter medium, its emission properties are directly addressed and evaluated. This can be beneficial in situations, where sensor contact to the emitter is not conceivable. Accordingly, no additional extra optical information transfer link for the readout and no separate temperature sensors for temperature monitoring are needed.

\smallskip

At the example of a semiconductor gain medium under laser operation, we demonstrate the analysis of its temperature by detection and evaluation of its specific spectral signal changes.
In this context, such a 'thermal fingerprint' represented by the direct and clear dependence of a semiconductor disk laser’s spectral signature on the device's temperature is used in this work for the training of multilayer feed-forward Neural Network (NN) models towards a temperature sensor based on the laser's signal. 
Because the emitter system is usually thermally coupled to its environment, such as the substrate or carrier, which in most cases is considered as the heatsink's body, or the local part of a photonic chip etc., it can be considered as a sensing device.
The direct analysis of collected laser light from the ’optical sensor’ device can be used for long-time temperature observation of far away lasers or remote objects that carry a laser, for which the spectral shift arises due to temperature differences with regard to the device’s reference temperature.
Our concept enables a non contact and long-distance examination of a gain-medium temperature and therewith a pathway to semiconductor-technology based miniaturized sensor elements, which for instance can be remotely probed by merely a simple miniaturized spectrometer coupled to a machine learning (ML) model.

\smallskip

Here, we investigated two lasers, which emit in the NIR spectral range at around $\SI{800}{}$ and $\SI{1070}{\nano\metre}$, respectively, with a compact NN architecture after hyperparameter optimization. Particularly, the detection was pursued with two cost-effective miniature spectrometers, which span spectrally-different wavelength regions between $\SI{400}{}$ and $\SI{1700}{\nano\metre}$. In this work, we demonstrate the temperature deduction performance for our most performant example configuration.

\section{Experiment and Computational Methods}

In order to acquire sufficiently enough data to train a NN, three spectrometers of similar type are employed for capturing labelled spectra at the same time - by that providing a connection between the three equally labelled captures. 
With the help of three similar yet different spectrometers owing to their different wavelength ranges, signal to noise ratios and their manufacturing-related deviations, it is possible to provide the network with slightly different spectra to prevent the network from overfitting and increase the robustness of our model.
These spectrometers employed were deemed suitable to capture the laser signal due to their device characteristics. Therefore, the chosen spectrometer should have an appropriate resolving power of the emission line and a high signal to noise ratio.
Moreover, concerning machine learning tasks, they should enable facile software integration and rapid data acquisition capacity for lab-task-tailored big data collection and processing.
To fulfil these requirements, spectrometers of the type \textit{PEBBLE} from \textit{IBSEN Photonics} \cite{ibsen} are chosen.

\smallskip

\begin{figure}[htp]
     \centering
     \includegraphics[width=\textwidth]{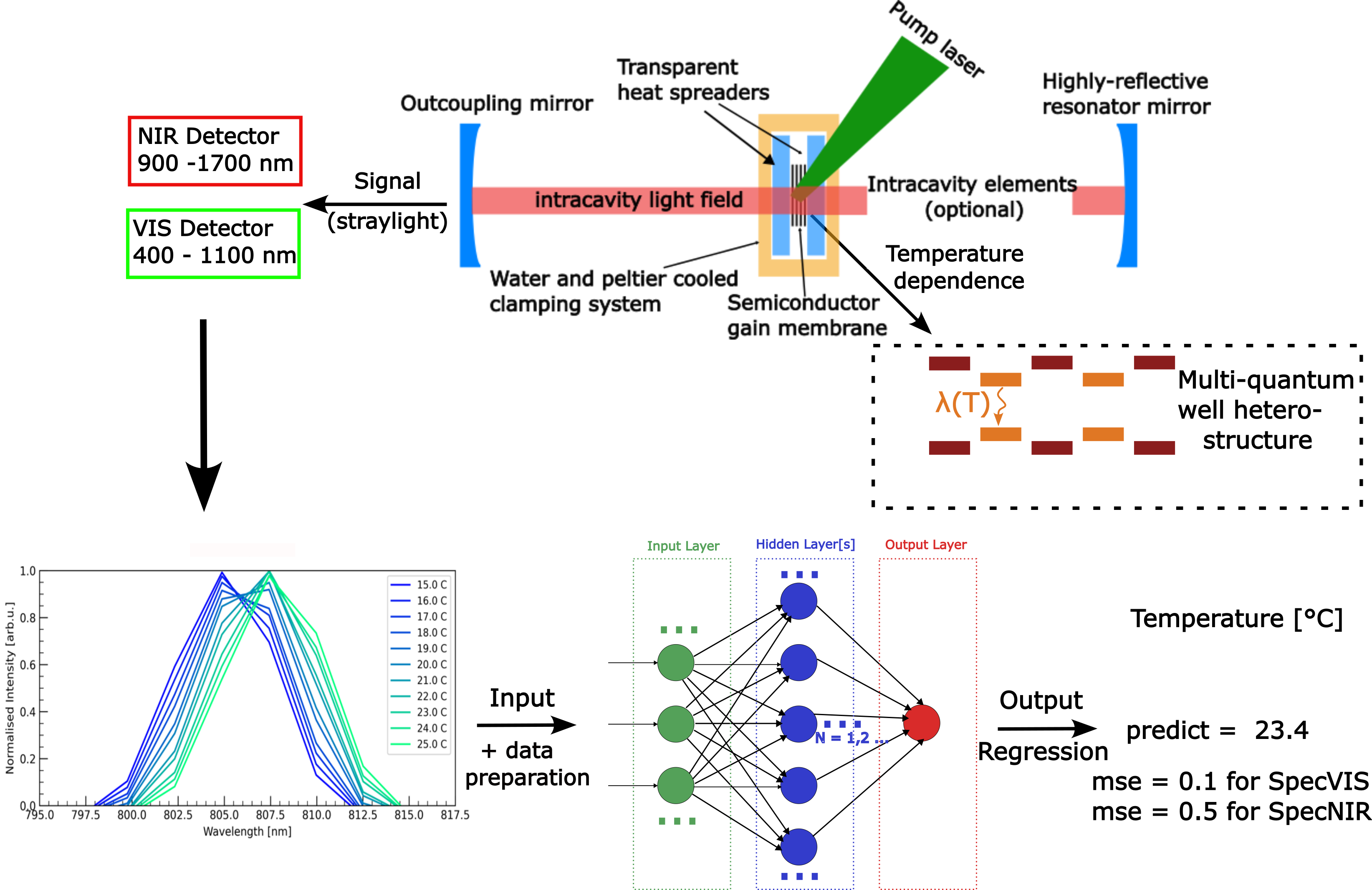}    
    \caption{ (a) Sketch of the working principle. The laser signal from our employed semiconductor device is captured by spectrometers, two for the VIS and one for NIR spectral range. The captured spectra are prepared for insertion into the NN when matching the input layer structure. After training with labelled data, the few-layer NN model infers the temperature owing to a supervised learning technique. For a given input spectrum, a temperature value is deduced.}
    \label{fig_1a_sketch}
\end{figure}

Figure \ref{fig_1a_sketch} sketches the experimental setup for the ML process principle for temperature regression by a trained NN. The emission spectra of the laser system are captured by the spectrometer and are labelled with their corresponding temperature according to the digital read out of the respective heat sink temperature. Afterwards, the prepared spectra are delivered for use as the input to the NN. In the training process, the regressor-type NN model learns to map the input values to a temperature as part of a supervised learning setting for the prediction task.
After the learning process, the network can predict the temperature merely based on a later provided input spectrum.


\begin{figure}[htp]
    \centering
    \includegraphics[width=\textwidth]{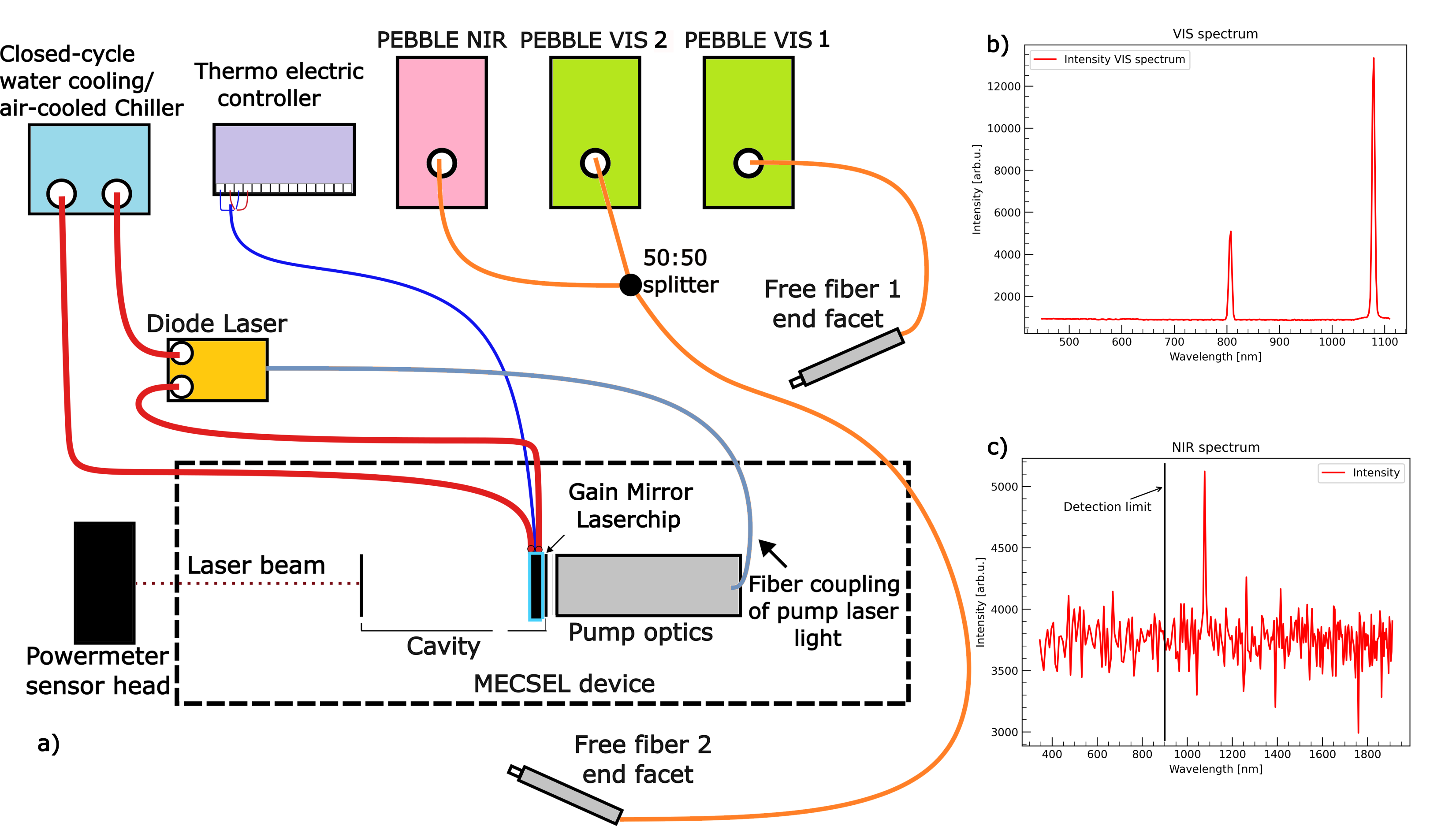}
    \caption{a) Sketch of the measurement arrangement for the acquisition of the laser system's spectral signature. Laser light is detected after back-scattering in free-space from an arbitrary screen (here power meter sensor head) by the fiber-coupled (orange lines) mini-spectrometers' detector for individual temperature settings. Here, the close-cycle water cooling (tubes drawn in red) and thermo-electric cooling device (wiring drawn in blue) act simultaneously for temperature control of the membrane gain chip, both adjusted and monitored software-wise. b)VIS raw spectra as recorded after the 50:50 (\%) splitter on the respective fiber-coupled spectrometer (fiber 2). c) NIR raw spectra as recorded after the same beam splitter on the respective fiber-coupled spectrometer (NIR). The minimum detectable wavelength is indicated by a vertical black line at $\SI{900}{\nano\meter}$.}
    \label{fig_2a_setup for acquisition}
\end{figure}

In this aforementioned setup there are two crucial issues for the task. Firstly, the proper spectra acquisition with the correct label, and secondly, the preparation of the data for the NN.
Figure \ref{fig_2a_setup for acquisition} with its more detailed representation of the data acquisition configuration addresses the first point. Note that this sketch shows quite accurately where the fibers, laser, powermeter sensor head and so forth are placed during the measurement, using the optical setup's geometrical arrangement, and locates core items -- other than the peripherical tools -- properly and in proportion to the cm positions on the table. However, the actual placements were done arbitrarily and, accordingly, the arrangement could have been chosen differently as well.

Both visible-light spectrometers  depicted (\textit{PEBBLE} "VIS 1" and "VIS 2") cover a wavelength range between $\SI{400}{} - \SI{1100}{\nano\meter}$ with a typical spectral resolution of $\SI{8}{\nano\meter}$ for the incorporated entrance slit width and type of transmission grating. The line-array detector in these spectrometers is a \textit{Hamamatsu S14739} \cite{pebble_vis_nir}.
The near infrared spectrometer (\textit{PEBBLE} "NIR") covers a spectral range between $\SI{900}{} - \SI{1700}{\nano\meter}$ with a specified spectral resolution of approximately $\SI{12}{\nano\meter}$ owing to the given set of incorporated transmission grating and line-array detector for infrared signals.
This type of NIR spectrometer has a \textit{Hamamatsu G13913} detector \cite{pebble_nir}.
Fiber 1 is connected to VIS \textit{PEBBLE} 1 and points from behind the Membrane External Cavity Surface Emitting Laser (MECSEL) device \cite{mexl} towards the overall open-cavity optically-pumped laser system and thus effectively collects stray light originating from both the pump laser and the MECSEL.
Fiber 2 similarly does, though at a different ratio in peak intensities, and is connected to a fiber-type $50:50$ splitter, which guides the light to both the VIS \textit{PEBBLE} 2 and the NIR \textit{PEBBLE}.
Due to the focus on MECSEL signal, and because the NIR \textit{PEBBLE} does not detect pump-laser light, fiber 2 is located rather next to the laser system, but pointing also in the direction of the laser beam blocked by the power-meter sensor head to collect stray light predominantly from the outcoupled MECSEL beam after a long pass filter, which attenuates the pump laser light in the laser's external beam path.
Nonetheless, the positions for both fiber tips (detecting facets) is chosen arbitrarily with the aim to record spectra with meaningful signal strength variations for the laser system.
Two example spectra are shown in Figure \ref{fig_2a_setup for acquisition} b) and Figure \ref{fig_2a_setup for acquisition} c) for the VIS spectrometer and NIR spectrometer respectively. The black vertical line shown in the NIR spectrum indicates the detection limit of that spectrometer.

\smallskip

The temperature of the pump laser is controlled by the water cooling/chiller system (model \textit{MRC150/300}, \textit{Laird Thermal Systems}\cite{mrc150_300}).
The temperature of the MECSEL is controlled by the \textit{thermo electric controller} (model \textit{1091}, \textit{Meerstetter engineering} \cite{tec_controller}).
Both temperatures can be set independently. 
For a given temperature range (here: $\SI{15}{\degreeCelsius} -  \SI{25}{\degreeCelsius}$) the setup can automatically sweep through the temperatures in pre-defined steps and capture an arbitrary number of spectra for each temperature.
Every spectrum recorded gets labelled with the corresponding temperature and the measured power of the laser. The laser power is measured with a power meter (model \textit{PM100-A}, Thorlabs \cite{powermeter}) in the beam path behind the outcoupler mirror and behind a long-pass filter (edge around $\SI{850}{\nano\metre}$).

\smallskip

For the training of the NN, spectra in the range from $\SI{15}{\degreeCelsius}$ to $\SI{25}{\degreeCelsius}$ in steps of $\SI{0.2}{\degreeCelsius}$ are recorded.
In addition to the recorded data, auxiliary data are generated which incorporate shifted spectra (here shifted by $\approx \SI{20}{\nano\metre}$) to ensure that the absolute wavelength of the laser system does not matter during the testing and also for uncoupling the laser line's spectral position/signature from the detector pixel location. Note that the shift is applied to both laser lines accordingly.
A corresponding noise part of the spectrum is copy-pasted to fill the side and complete the spectral range of the captures to match that of original data.
Real spectra and auxiliary data combined compose a set of approximately 268.000 spectra. However, only the 134.000 captured original spectra are used to train and test the model, whereas the shifted spectra are used to validate the model afterwards.
The table in the results section with mean squared error (mse) listings refers to the validation with these shifted spectra. For a better comparison, a reference table of the mse during the (self)-training is given in the appendix in Tables \ref{tab:test_results} and \ref{tab:test_results_without_pump}, respectively.  

\begin{figure}[h!]
     \centering
     \begin{subfigure}[c]{0.5\textwidth}
         \centering
         \includegraphics[width=\textwidth]{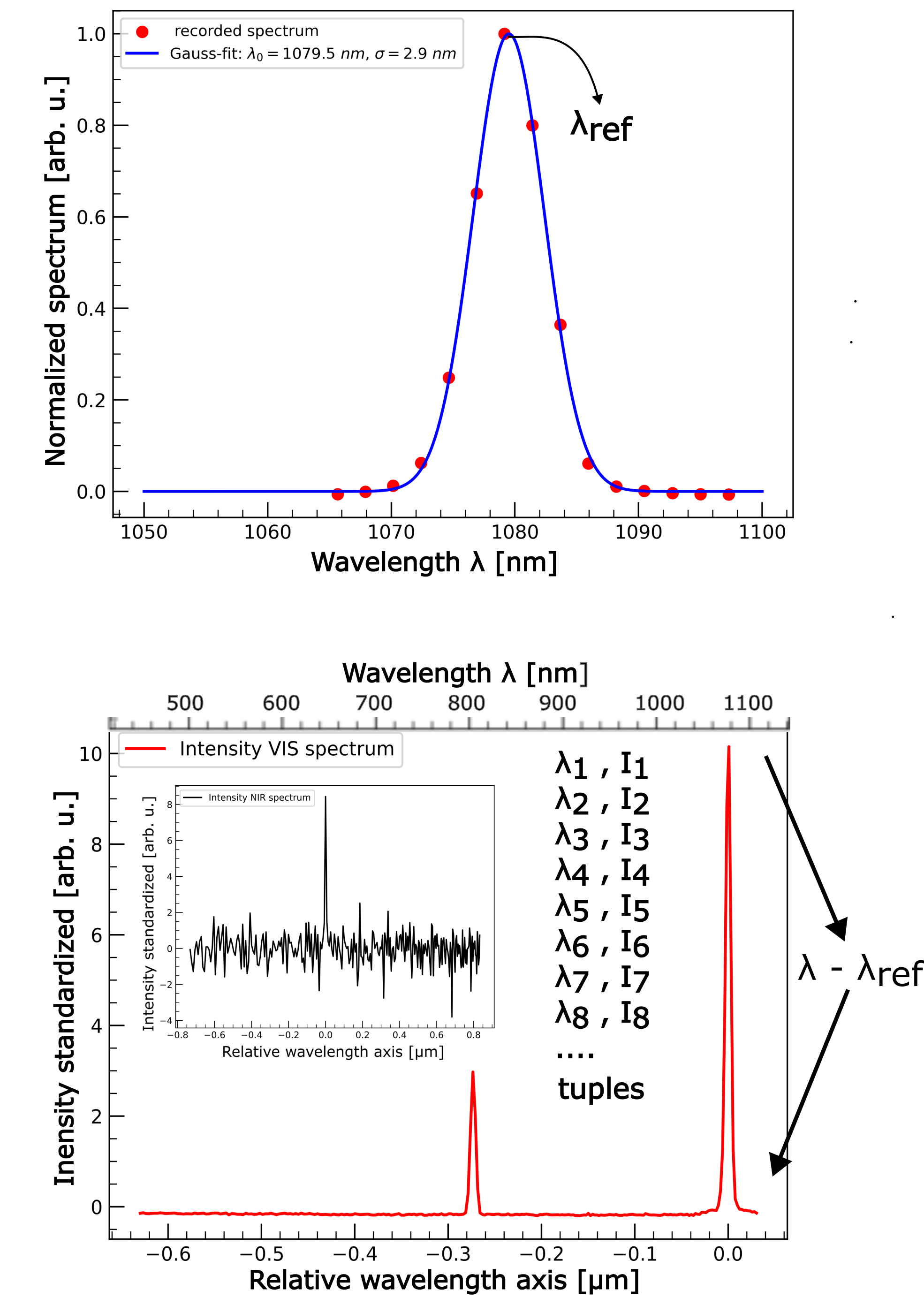}
         \caption{The reference wavelength is found by a Gaussian fit to the obtained spectra at the chosen reference temperature. Afterwards, the relative wavelength axis is generated. The obtained spectrum with the rescaled intensities exhibits the reference wavelength signal at 0. These data. i.e. the tuples of (relative) wavelength and intensity, become the input of the used ML model.}
         \label{fig_3a_data_prep}
     \end{subfigure}
     \hfill
     \begin{subfigure}[c]{0.4\textwidth}
         \resizebox{6cm}{!}{
         \begin{tabular}{c} \hline
          \thead{Input Layer, input: [(None, 256, 2)] \\
          Input Layer, output: [(None, 256,2)]}  \\ \hline
          $\downarrow $ \\ \hline
          \thead{Flatten Layer [(None, 256, 2)] \\
          Flatten, output: [(None, 512)]} \\ \hline
          $\downarrow $ \\ \hline
          \thead{Hidden Layer 1: [(None, 512)] \\
          Dense, output: [(None, 256)]} \\ \hline
          $\downarrow$ \\ \hline
          \thead{Hidden Layer 2, input: [(None, 256)] \\
          Dense, output: [(None, 256)]} \\ \hline
          $\downarrow$ \\ \hline
          \thead{Hidden Layer 3, input: [(None, 256)] \\
          Dense, output: [(None, 256)]}\\ \hline
          $\downarrow$ \\ \hline
         \thead{Output Layer 1, input: [(None, 256)] \\
          Dense, output: [(None, 1)]} \\ \hline
  \end{tabular}
  }
  \caption{ Structure of the deployed 5 layer
NN. The input layer is followed by a layer to flatten the 2 dimensional(D) input into a 1-D vector with alternating wavelength and intensity values (of type float numbers). The following three layers are fully-connected (dense) Hidden layers with 256 units in each layer. The output layer consists of a single unit for the output of the temperature (float).}
     \end{subfigure}
\end{figure}

To address the second crucial issue, that is, of preparing the data, Figure \ref{fig_3a_data_prep} indicates the preparation steps. In order to train the network efficiently, all spectra undergo the same preparation.
Since a temperature change induces a shift in the emission signal spectrally, the absolute wavelength is not suitable for the inference of the temperature out of the given signal, as this can lead to the effect that in the sensor-deployment case an originally slightly de-tuned laser is interpreted as a shift of temperature, even though the change originates e.g from design, production or other technical differences for the same device category/chip type (thermal dependencies).
To account for this, a reference spectrum at a chosen temperature, in this case  $\SI{20}{\degreeCelsius}$, is generated. 
The wavelength axis is then transformed to a relative spectrum using a central
wavelength obtained from the reference spectrum. The reference value can for example be found by extracting the central emission wavelength with a Gaussian fit to the reference spectrum as depicted in Figure \ref{fig_3a_data_prep}. 
The relative wavelength axis is then obtained according to the simple expression $\delta \lambda = \lambda - \lambda_{ref}$ where $\lambda$ is the measured wavelength and $\lambda_{ref}$ the reference value.
The signal axis (i.e. the intensity counts) will then be rescaled such that its values have a mean of zero and a standard deviation of one. This ensures that all features have the same order of magnitude and, thus, the NN can fit the weights properly \cite{Bishop1994NeuralNA}.
Afterwards, the counts and the corresponding wavelengths get combined so that one obtains 256 tuples ($l_{i}/\lambda_{i}$ with counts and wavelength for each 256-entry long spectrum with $i$ as pixel counter, i.e. array position).
This 2D structure is the input to the NN. The first layer of the NN is a flatten layer to flatten the 2D structure of the input to a 1D vector with alternating wavelength and intensity entries.

\section{Results and Discussion}
For the training process a 5 layer Deep Neural Network with 3 hidden layers is used.
The 5 layer NN was chosen through a \textit{grid-search} from the python \textit{scikit} module. Attention has been paid to both accuracy and computational resources, so that the 5 layer network appears more potent compared to shallower NNs of the same type but at the same time it is good enough that deeper architectures and complexity of the model appear unfavourable and not profitable, as the precision gain for more then 5 layer is low compared to the extra cost in computational effort.
In addition, suitable hyperparameters are found with the same \textit{grid-search}. In this work, the \textit{sigmoid} activation function, 256 units in each hidden layer, 25 epochs (cycles of error minimization with Adam optimizer \cite{Kingma2014AdamAM}) of learning and a batch size of 50 are used. Note that these parameters are only the best within the performed grid search and there might be even better combinations that were not checked here. For the training procedure, the originally recorded and input-prepared data set of 134.000 spectra is split into $\SI{70}{\percent}$ for training and $\SI{30}{\percent}$ for testing during the learning phase. The whole $\SI{20}{\nano\metre}$ shifted auxiliary data set of 134.000 spectra is used solely for (post-learning) performance testing, i.e. validation. The \textit{mean-squared-errors (mse)} for different laser--spectrometer pairs in the validation process are summarized in Table \ref{tab:results}. The left side of the table shows the \textit{mse} when the model was trained with spectra from the VIS spectrometer only. On the right side, VIS and NIR spectra were used to train the model. The row "Test Data" shows which kind of spectra were used for testing or validation e.g. from one or two different spectrometers ("NIR", "VIS1", "VIS2", or both VIS devices, short "VIS"). The three different step sizes (table rows) show the impact on the inference result when the temperature steps in sweeps are differently coarse for data used in the training and validation process. For comparison, the \textit{mse} during the (self)-testing in the learning phase for the case "VIS + NIR" learning data and "VIS" testing data reached values as low as $\approx 0.06$, which is similar to the \textit{mse} obtained from the presented validation step. The validation solely relies on the auxiliary data with $\SI{20}{\nano\metre}$ shifted spectra after input preparation, without retraining the NN to the previously "unseen" shifted spectra. For comparison purposes, the \textit{mse} obtained as a result of the learning procedure after the last training epoch is summarized in Table \ref{tab:test_results} (Appendix).

\begin{table}[h!]
\centering
\begin{tabular}{c|cccc|cccc}
Learning Data      & \multicolumn{4}{c}{VIS}& \multicolumn{4}{c}{VIS + NIR} \\ \hline
Test Data  & NIR   & VIS   & VIS 1  & VIS 2    & NIR   & VIS   & VIS 1  & VIS 2                        \\ \hline
\textit{mse} $\SI{0.2}{\degreeCelsius}$ steps & \cellcolor[HTML]{CB0000}9.41  & \cellcolor[HTML]{32CB00}0.04 & \cellcolor[HTML]{32CB00}0.05 & \cellcolor[HTML]{32CB00}0.04 & \cellcolor[HTML]{F8A102}2.83  & \cellcolor[HTML]{32CB00}0.04 & \cellcolor[HTML]{32CB00}0.04 & \cellcolor[HTML]{32CB00}0.04 \\
\textit{mse} $\SI{0.5}{\degreeCelsius}$ steps & \cellcolor[HTML]{CB0000}9.85  & \cellcolor[HTML]{32CB00}0.04 & \cellcolor[HTML]{32CB00}0.04 & \cellcolor[HTML]{32CB00}0.04 & \cellcolor[HTML]{F8A102}3.08  & \cellcolor[HTML]{32CB00}0.05 & \cellcolor[HTML]{32CB00}0.04 & \cellcolor[HTML]{32CB00}0.05 \\
\textit{mse} $\SI{1}{\degreeCelsius}$ steps   & \cellcolor[HTML]{CB0000}10.12  & \cellcolor[HTML]{32CB00}0.04 & \cellcolor[HTML]{32CB00}0.04 & \cellcolor[HTML]{32CB00}0.04 & \cellcolor[HTML]{F8A102}3.30 & \cellcolor[HTML]{32CB00}0.04 &\cellcolor[HTML]{32CB00}0.04 & \cellcolor[HTML]{32CB00}0.05
\end{tabular}%
\caption{Test for the 5-Layer Deep Network tested with a $\SI{20}{\nano\metre}$ shifted data set without retraining the network to the shifted spectra. The network is trained with the real recorded spectra only.}
\label{tab:results}
\end{table}

\smallskip

When trained only with VIS spectra, the model has problems inferring from the NIR \textit{PEBBLE} data, as expected. 
However, adding the NIR spectra to the training data improves the result significantly, when MECSEL data without pump line are part of the evaluation process.
The VIS spectra reach a mean-square-error below $\SI{0.1}{\degreeCelsius}$, while the NIR spectra remain in the range of $\approx \SI{3}{\degreeCelsius}$, in accordance to the influence of the spectrometers' resolving power, ie, the capability of the used spectrometer to resolve the small shifts of the emission wavelength in the captured spectra. This indicates that a suitable spectrometer--laser pair should be used and that a calibration of the spectrometer--laser system might be helpful.
In addition, the accuracy with the pump laser signal is higher compared to the MECSEL signal (as the pump laser outperforms the MECSEL in terms of linewidth). Accordingly, this results from the resolving power in combination with the linewidth of the laser lines. Because the linewidth of the pump laser is about $\SI{20}{\percent}$ narrower than the linewidth of the MECSEL, the trained NN can more accurate infer temperatures from the predominantly contributing pump laser signal; in simple words, the NN will likely focus on the line from which temperature-related behaviours are most obvious, the diode laser here. To test this hypothesis, a data set with artificially removed pump laser from the spectra is fed into the NN for the training, testing and validation of the network in the same way as before. The results of this check are given in Table \ref{tab:test_without_pump}. For a comparison to results directly obtained from the training procedure, a corresponding \textit{mse} overview is given in Table \ref{tab:test_results_without_pump}.

\begin{table}[h!]
\centering
\begin{tabular}{c|cccc|cccc}
Learning Data     & \multicolumn{4}{c}{VIS}& \multicolumn{4}{c}{VIS + NIR} \\ \hline
Test Data & NIR   & VIS   & VIS 1  & VIS 2    & NIR   & VIS   & VIS 1  & VIS 2                        \\ \hline
\textit{mse} $\SI{0.2}{\degreeCelsius}$ steps & \cellcolor[HTML]{CB0000}10.91  & \cellcolor[HTML]{32CB00}0.43 & \cellcolor[HTML]{32CB00}0.57 & \cellcolor[HTML]{32CB00}0.28 & \cellcolor[HTML]{F8A102}2.76 &\cellcolor[HTML]{32CB00}0.70 & \cellcolor[HTML]{F8A102}1.09 & \cellcolor[HTML]{32CB00}0.30 \\
\textit{mse} $\SI{0.5}{\degreeCelsius}$ steps & \cellcolor[HTML]{CB0000}11.36  & \cellcolor[HTML]{32CB00}0.50 & \cellcolor[HTML]{32CB00}0.67 & \cellcolor[HTML]{32CB00}0.33 & \cellcolor[HTML]{F8A102}3.06  & \cellcolor[HTML]{32CB00}0.75 & \cellcolor[HTML]{F8A102}1.14 & \cellcolor[HTML]{32CB00}0.36 \\
\textit{mse} $\SI{1}{\degreeCelsius}$ steps   & \cellcolor[HTML]{CB0000}11.75  & \cellcolor[HTML]{32CB00}0.51 & \cellcolor[HTML]{32CB00}0.68 & \cellcolor[HTML]{32CB00}0.33 &\cellcolor[HTML]{F8A102}3.28 & \cellcolor[HTML]{32CB00}0.77 & \cellcolor[HTML]{F8A102}1.15 & \cellcolor[HTML]{32CB00}0.39
\end{tabular}%

\caption{Test for the 5-Layer Deep Network trained with the same hyperparameters as above but this time the pump laser is artificially removed from the data. Afterwards, the relative shift was applied without retraining the network to the artificially altered spectra.}
\label{tab:test_without_pump}
\end{table}

Clearly,  the \textit{mse} become worse when excluding the pump laser line. This is understandable due to the fact that the pump laser line can be better resolved in the spectrometer, and thus, the shift can be detected more easily. Accordingly, temperature inference can become improved, or for broader lines worsened.
As the validation data consists of shifted spectra, it is visible that the relative wavelength approach overcomes the bottleneck of the recorded "absolute" wavelength and ensures a temperature inference independent of the absolute wavelength. This enables a more versatile prediction of the temperature for a given spectrometer--laser pair configuration without transfer learning involved, e.g. accounting for production tolerances.
Note that one can also train the network without the relative wavelength axis only with rescaled features. However, this limits the system to an absolute wavelength information. Nonetheless, when operating a stable laser system, where no major changes of the emission wavelengths (e.g. modejumps, production variations, power-induced shifts etc.) are expected, the inference of the temperature can be comparably precise as with the relative wavelength information. For a suitable laser--spectrometer pair with a reasonable resolving power of the laser emission line, in the case of involving the pump laser and absolute wavelength scale a \textit{mse} of $\SI{0.13}{\degreeCelsius}$ is found. This is slightly worse than the primarily discussed approach, but still accurate enough for a range of  possible use cases such as long time temperature monitoring where trends matter more than absolute temperatures. If the model is trained with heavier and more complex data sets featuring different laser signatures, e.g. with a removed pump laser or with data paying attention to power-dependent issues or other aspects of the signal properties, certain limitations could typically be overcome to some extent at the cost of computational load and data complexity. Correspondingly, in effect, here we highlight the advantage of a compact configuration and ressource effective approach.
Moreover, smaller networks with less layers or less units can be used. However, this comes with the risk of reducing a bit the precision, or leading to a tendency of underfitting data, but it may well enable resource-saving in cases where this might be critical and accuracy is not the most important consideration. With a network consisting of 3 layers and 256 neurons in each layer a \textit{mse} of $\approx \SI{0.14}{\degreeCelsius}$ is found when trained with VIS + NIR data and tested with data from VIS PEBBLE 1, i.e. with data that contains both the MECSEL and pump laser line. Compared to the 5-layer network the accuracy dropped slightly but the model is still able to infer the temperature up to the fist decimal point.

\smallskip

Instead of using a regressor network, one could also investigate the performance of a classifier network. In order to do so, one would need to group the temperatures into classes before training the network. This, however, has the limitation that the prediction (i.e. sensory) resolution is fixed as soon as the classes are fixed but enables to set the temperature resolution beforehand. Furthermore, long time laser signal observation could be utilized in future model developments to train such NNs to predict target property alterations, changes based on temperature changes and time-trace specialized network architectures could lead to effective monitoring tools regarding external influences on selected emitter properties.

\smallskip 

In addition to the here reported functionality, machine learning image-/video-analysis tools could be used to detect signal anomalies in time series.
Through the flexibility of the ML models, they can be adjusted for the given device parameters, configurations and their applications, including in precision with respect to computational needs as well as application requirements.
Besides applications in temperature sensing, such NN models can be used to investigate and infer other laser parameters (cf. recent examples in the field of laser examinations/ultrafast photonics by e.g. D. Zibar et al. \cite{Darko} and G. Genty et al. \cite{review-ml-photonics}, to name but a few), such as pulse duration or cavity parameters from spectral signatures of a laser system, benefiting from transfer learning capabilities for pre-trained models in combination with suitable and sufficiently large data sets.

\section{Conclusion}

We demonstrated a compact NN system with the capability to infer the temperature of two semiconductor lasers by a machine-learning method based on analysis of their recorded spectral signature. In combination with miniature spectrometers in the VIS and NIR spectral range with sufficiently high resolving power concerning the signal of interest, the accuracy reached as good as about $\SI{0.05}{\degreeCelsius}$. This can be precise enough for possible remote/contactless sensing applications or for miniaturized emitter-based sensor elements with micro-spectrometers.
This concept enables a non-contact and long-distance examination of a gain-medium temperature and therewith a pathway to semiconductor-technology based miniaturized sensor elements, which for instance can be remotely probed by merely a simple spectrometer coupled to a machine learning model. 

\smallskip

Moreover, RNN designed to remember sequences of inputs over longer periods can help to perform prediction tasks over longer periods, thereby enlarging the use cases of our study due to its ability to recognize temporal evolution. By capturing temperature temporal dependencies, real-time application of the machine learning model with signals/systems varying dynamically over time could be addressed effectively. Combined with the ability of CNNs to extract spatial features, spectral monitoring could be done with known correlations between sensor signals and different environmental impacts, improving the ability to differentiate between multiple wavelength shift causes sufficiently for a given application scenario.

\medskip

\smallskip
\textbf{Acknowledgements} \par 
Financial support by the Deutsche Forschungsgemeinschaft (German Research Foundation, DFG: RA2841/12-1)---Projektnummer 456700276---is acknowledged.

\smallskip
\textbf{Conflict of Interest} \par 
The authors declare no conflict of interest.

\smallskip
\textbf{Author Contributions} \par 
A.R.-I. conceived the experiment and guided J.M. on the machine learning project. Data acquisition and preparation was conducted by J.M. The multilayer-neural-network model for the sensing application was jointly realized by J.M. and A.R.-I. The supervised-learning model configurations and their performance were examined by J.M. The results were discussed and summarized in a manuscript by both authors.

\smallskip

\bibliographystyle{MSP}
\bibliography{Paper_arash_jakob.bib}

\appendix

\section{Comparison tables for the mean-squared-error values}

For comparison to the validation tables \ref{tab:results}, \ref{tab:test_without_pump} given in the results section, here the \textit{mse} during the self testing at the end of the learning process is provided.
The hyperparameters (epochs, optimizer, batch size etc) of the model as well as the data input preparation (rescalings) are the same.

\begin{table}[h!]
\centering
\begin{tabular}{c|cccc|cccc}
Learning Data      & \multicolumn{4}{c}{VIS} & \multicolumn{4}{c}{VIS + NIR}  \\ \hline
Test Data & NIR   & VIS  & VIS 1  & VIS 2  & NIR    & VIS    & VIS 1   & VIS 2  \\ \hline
\textit{mse} $\SI{0.2}{\degreeCelsius}$ steps & \cellcolor[HTML]{CB0000}9.47  & \cellcolor[HTML]{32CB00}0.07 & \cellcolor[HTML]{32CB00}0.06 & \cellcolor[HTML]{32CB00}0.08 & \cellcolor[HTML]{F8A102}3.26 & \cellcolor[HTML]{32CB00}0.07 &\cellcolor[HTML]{32CB00}0.06 & \cellcolor[HTML]{32CB00}0.08 \\
\textit{mse} $\SI{0.5}{\degreeCelsius}$ steps & \cellcolor[HTML]{CB0000}9.89  & \cellcolor[HTML]{32CB00}0.07 & \cellcolor[HTML]{32CB00}0.06 & \cellcolor[HTML]{32CB00}0.08 &\cellcolor[HTML]{F8A102}3.60  &\cellcolor[HTML]{32CB00}0.07 & \cellcolor[HTML]{32CB00}0.06 & \cellcolor[HTML]{32CB00}0.09 \\
\textit{mse} $\SI{1}{\degreeCelsius}$ steps   & \cellcolor[HTML]{CB0000}10.16  & \cellcolor[HTML]{32CB00}0.07 & \cellcolor[HTML]{32CB00}0.06 & \cellcolor[HTML]{32CB00}0.09 & \cellcolor[HTML]{F8A102}3.88 & \cellcolor[HTML]{32CB00}0.07 & \cellcolor[HTML]{32CB00}0.05 & \cellcolor[HTML]{32CB00}0.09
\end{tabular}%
\caption{Results of the 5-Layer Deep Network trained with the original recorded data prepared as explained in the main text. These \textit{mse} were achieved by the model after the end of the learning process before the validation with the shifted data.}
\label{tab:test_results}
\end{table}

\begin{table}[h!]
\centering
\begin{tabular}{c|cccc|cccc}
Learning Data & \multicolumn{4}{c}{VIS}   & \multicolumn{4}{c}{VIS + NIR}   \\ \hline
Test Data & NIR      & VIS  & VIS 1   & VIS 2    & NIR   & VIS   & VIS 1   & VIS 2   \\ \hline
\textit{mse} $\SI{0.2}{\degreeCelsius}$ steps & \cellcolor[HTML]{CB0000}10.74  & \cellcolor[HTML]{32CB00}0.68 & \cellcolor[HTML]{32CB00}0.90 & \cellcolor[HTML]{32CB00}0.46 & \cellcolor[HTML]{F8A102}3.22 & \cellcolor[HTML]{32CB00}0.91 & \cellcolor[HTML]{F8A102}1.34 & \cellcolor[HTML]{32CB00}0.47 \\
\textit{mse} $\SI{0.5}{\degreeCelsius}$ steps & \cellcolor[HTML]{CB0000}11.28  & \cellcolor[HTML]{32CB00}0.77 & \cellcolor[HTML]{32CB00}1.07 & \cellcolor[HTML]{32CB00}0.48 &\cellcolor[HTML]{F8A102}3.63  & \cellcolor[HTML]{32CB00}0.97 & \cellcolor[HTML]{F8A102}1.41 & \cellcolor[HTML]{32CB00}0.53 \\
\textit{mse} $\SI{1}{\degreeCelsius}$ steps   & \cellcolor[HTML]{CB0000}11.64 & \cellcolor[HTML]{32CB00}0.86 & \cellcolor[HTML]{32CB00}1.22 & \cellcolor[HTML]{32CB00}0.50 &\cellcolor[HTML]{F8A102}3.89 & \cellcolor[HTML]{F8A102}1.05 & \cellcolor[HTML]{F8A102}1.51 &\cellcolor[HTML]{32CB00}0.59
\end{tabular}%

\caption{Results 5-Layer Deep Network trained with the data set containing only spectra without the pump laser. The data \textit{mse} values were taken again from the last iteration of the learning process.}
\label{tab:test_results_without_pump}
\end{table}

When comparing these values with the validation values, it can be seen that the validation \textit{mse} is even below the one for the in-training testing. This most likely originates from the fact that the validation data set with 134.000 auxiliary spectra contains significantly more data points than the in-training test set, which is made up of only $\SI{30}{\percent}$ of the 134.000 (original) spectra. Thus, the spread reduces and high \textit{mse} values tend to lose their impact resulting in a lower \textit{mse}.

\end{document}